\documentclass[a4paper,pra,twocolumn,superscriptaddress]{revtex4}

\usepackage[english]{babel}
\usepackage[utf8]{inputenc}
\usepackage{graphicx,epsfig,color}
\usepackage{latexsym}
\usepackage{amsfonts} 
\usepackage{amsmath}
\usepackage{textcomp}
\usepackage{hyperref}
\usepackage{placeins}

\def\ket|#1>{| #1 \rangle}
\def\bra<#1|{\langle #1 |}
\def\<{\langle}
\def\>{\rangle}
\def\{{\lbrace}
\def\}{\rbrace}
\def\({\left(}
\def\){\right)}
\def\beq{\begin{equation}}
\def\eeq{\end{equation}}
\def\eff{\mathrm{eff}}
\def\Z{\mathbb{Z}}

\begin{document}

\title{Power Accretion in Social Systems}%

\author{Silvia N. Santalla}
\affiliation{Dto. F\'{\i}sica and Grupo Interdisciplinar de Sistemas
  Complejos, Universidad Carlos III de Madrid, Spain}

\author{Kostadin Koroutchev}
\affiliation{Escuela Polit\'ecnica Superior, Universidad Aut\'onoma de
Madrid, Cantoblanco, Spain}

\author{Elka Korutcheva}
\affiliation{Dto. F\'{\i}sica Fundamental,
  Universidad Nacional de Educaci\'on a Distancia (UNED), Spain}
\affiliation{Dep. Theoretical Physics, G. Nadjakov Institute of Solid
  State Physics, Bulgarian Academy of Sciences, 72 Tzarigradsko
  Shaussee Blvd.  1784 Sofia, Bulgaria}

\author{Javier Rodr\'{\i}guez-Laguna}
\affiliation{Dto. F\'{\i}sica Fundamental, Universidad Nacional de
  Educaci\'on a Distancia (UNED), Spain}

\begin{abstract}
  We consider a model of power distribution in a social system where a
  set of agents play a simple game on a graph: the probability of
  winning each round is proportional to the agent's current power, and
  the winner gets more power as a result. We show that, when the
  agents are distributed on simple 1D and 2D networks, inequality
  grows naturally up to a certain stationary value characterized by a
  clear division between a higher and a lower class of agents. High
  class agents are separated by one or several lower class agents
  which serve as a geometrical barrier preventing further flow of
  power between them. Moreover, we consider the effect of
  redistributive mechanisms, such as proportional (non-progressive)
  taxation. Sufficient taxation will induce a sharp transition towards
  a more equal society, and we argue that the critical taxation level
  is uniquely determined by the system geometry. Interestingly, we
  find that the roughness and Shannon entropy of the power
  distributions are a very useful complement to the standard measures
  of inequality, such as the Gini index and the Lorenz curve.
\end{abstract}

\date{July 1, 2019}

\maketitle

\section{Introduction}

Inequality of income, wealth or power is one of the main social and
political concerns \cite{Hurst13,Piketty14}, since it undermines
social welfare, democracy \cite{Gilens14} and economic growth
\cite{Milanovic18}. Thus, a correct understanding of the dynamics of
inequality has become one of the main research topics of theoretical
social science and economics. There are different theoretical
approaches in order to explain social inequality
\cite{Castellano09,Pluchino18}. On one hand there are models based on
the disparity of human abilities, which claim that inequality
increases are related to a large extent to the growth in technological
complexity, that puts a bigger prize on certain rare skills. On the
other hand there are theoretical approaches based on self-organization
and the intrinsic instability associated to the accretion of wealth:
{\em money begets money}. In addition, there are institutional and
political factors, such as taxation or other governmental policies,
which bear a strong influence on social inequality
\cite{Hurst13,Piketty14,Warwick19}.

Statistical mechanics can play a significant role in this
endeavour. The interchange of wealth between individuals was compared
to the interchange of energy between the molecules of a gas which
leads to the Boltzmann-Gibbs distribution. Yet, in 1960 Mandelbrot
\cite{Mandelbrot60} remarked the difficulty of making this view
compatible with one of the main empirical observations about the
distribution of income or wealth, known as {\em Pareto's law}: the
population with income above $u$ falls like $u^{-\alpha}$ for large
enough $u$. In 1996, Stanley et al. \cite{Stanley96} remarked that
power-laws were, in fact, ubiquitous in physical systems with
long-range correlations. Moreover, they coined the term {\em
  econophysics}, in analogy to {\em biophysics}, to describe the
application of statistical mechanical concepts and methods to the
study of the economy \cite{Schinkus13}, with {\em agents} playing the
role of atoms \cite{Gupta06,Martino06,Chakraborti11,Galam12}.

These agents can be simple or complex, and in modern approaches they
are even allowed to learn from their experience \cite{Pinasco18}. Most
analysis in econophysics favor agents following simple rules, yet
showing rich dynamics. In 2000, Bouchaud and M\'ezard showed that a
simple model with random speculative trading might be mapped to the
well-known problem of directed polymers \cite{Bouchaud00}, showing a
rather sharp transition between a relatively egalitarian and an
extremely unequal phase. The same year, Dr\u agulescu and Yakovenko
\cite{Dragulescu00}, and Chakraborti and Chakrabarti
\cite{Chakraborti00}, used simple models with conserved wealth and
random interchanges, with or without savings, to describe different
equilibrium distributions. These models were found to yield Pareto
distributions in some regimes
\cite{Patriarca04,Cordier05,Duering08,Calbet11,Katriel14,Boghosian14a,Boghosian14b},
see \cite{Yak09,Chakrabarti13,Chatterjee15} for reviews. Some recent
models have considered the effect of personal savings and taxation
\cite{Berman16}. Other studies have focused on the appearance of
social classes \cite{Burda18}, the extraordinary velocity of growth of
the top earners \cite{Gabaix} or how inequality may induce economic
crisis without requiring external shocks \cite{Benisty}. Many of these
models are built on multiplicative stochastic processes, for which the
approach to equilibrium can be extremely slow, and the validity of the
ergodic hypothesis is questionable
\cite{LZ1,LZ2,Benisty,Berman17,Stojkovski19}. Interestingly, recent
work shows that pooling and sharing of resources (i.e. redistribution)
may increase the growth rate \cite{Peters19,Stojkovski19}.

We propose an extremely simple model combining some features which are
already present in the literature: (a) agent interactions that amplify
inequalities: the more you earn, the easier it is to earn even more;
this active element of wealth motivates us to call it {\em power}; (b)
geometric constraints: agents can only interact with their neighbors;
(c) a global redistributive mechanism, which we will call {\em
  taxation}. In addition we have explored the above features in a {\em
  graph} and have quantified the wealth inequality using both standard
measures and statistical mechanical concepts which are not usual in
the econophysics context, such as the roughness and the Shannon
entropy. As we will show, our proposed dynamics gives rise generically
to a higher and a lower class of agents, created through the
amplification of initial random fluctuations in combination with
geometrical constraints. The stationary state is strongly dependent on
the early history of the system, and ergodicity is broken when
taxation is absent. Low amounts of taxation can have subtle
counter-intuitive consequences. Yet, high enough tax levels lead to a
phase transition towards a much more equal system, where ergodicity is
restored.

Similar phenomena of amplification of random fluctuations make
appearance in other areas. For example, many models of interfacial
dynamics show a higher growth rate at peaks than at valleys, giving
rise to the so-called {\em shadowing instability}
\cite{Yao.93,Krug.96}, which explains e.g. the characteristic
flower-like shapes of bacterial colonies in a medium with limited
nutrients \cite{Santalla.18,Santalla.18b}. Moreover, fluctuations are
also amplified in {\em P\'olya's urn model} \cite{Polya30,Mahmoud09},
where we are asked to pick a ball from an urn and replace it with
several balls of the same color. Interestingly, P\'olya's urn model
has found several applications in social science, e.g. to innovation
\cite{Loreto17}.

We would like to emphasize that our model is built on two elements
which have been widely employed in the statistical mechanics and the
econophysics literature: random interchanges leading to unequal
distributions and a smoothing mechanism. Our focus, nonetheless, will
be on their interaction {\em through a graph} and the geometrical
constraints imposed on the growth of inequality. The aim of this work
is merely to present an extremely simple statistical mechanical model
whose merit is to characterize how the amplification of noisy events
can give rise to the creation of a strong class division, and some
efficient ways through which these effects might be mitigated. We do
not put forward any claims regarding actual social inequality.

The article is organized as follows. Section \ref{sec:powergame}
discusses the {\em power game} in some detail. The case of two players
is exposed in Sec. \ref{sec:twoplayers}. In Sec. \ref{sec:1d} we
consider a one-dimensional array of players in detail, combining tools
from economics, information theory and kinetic roughening. Other graph
structures are discussed in Sec. \ref{sec:2d}, clarifying the nature
of the transition. The article ends with conclusions and some
suggestions for further work.


\section{The Power Game}
\label{sec:powergame}

Let us consider $N$ agents, connected through a certain graph ${\cal
  G}$. Agent $i$ is endowed with a certain magnitude which we will
call {\em power}, $w_i\geq 0$. Total power is normalized to be one,

\beq
\sum_{i=1}^N w_i=1,
\label{eq:norm}
\eeq
The initial distribution of power will always be homogeneous,
i.e. $w_i=1/N$ for all $i$.

\begin{figure}
  \includegraphics[width=8cm]{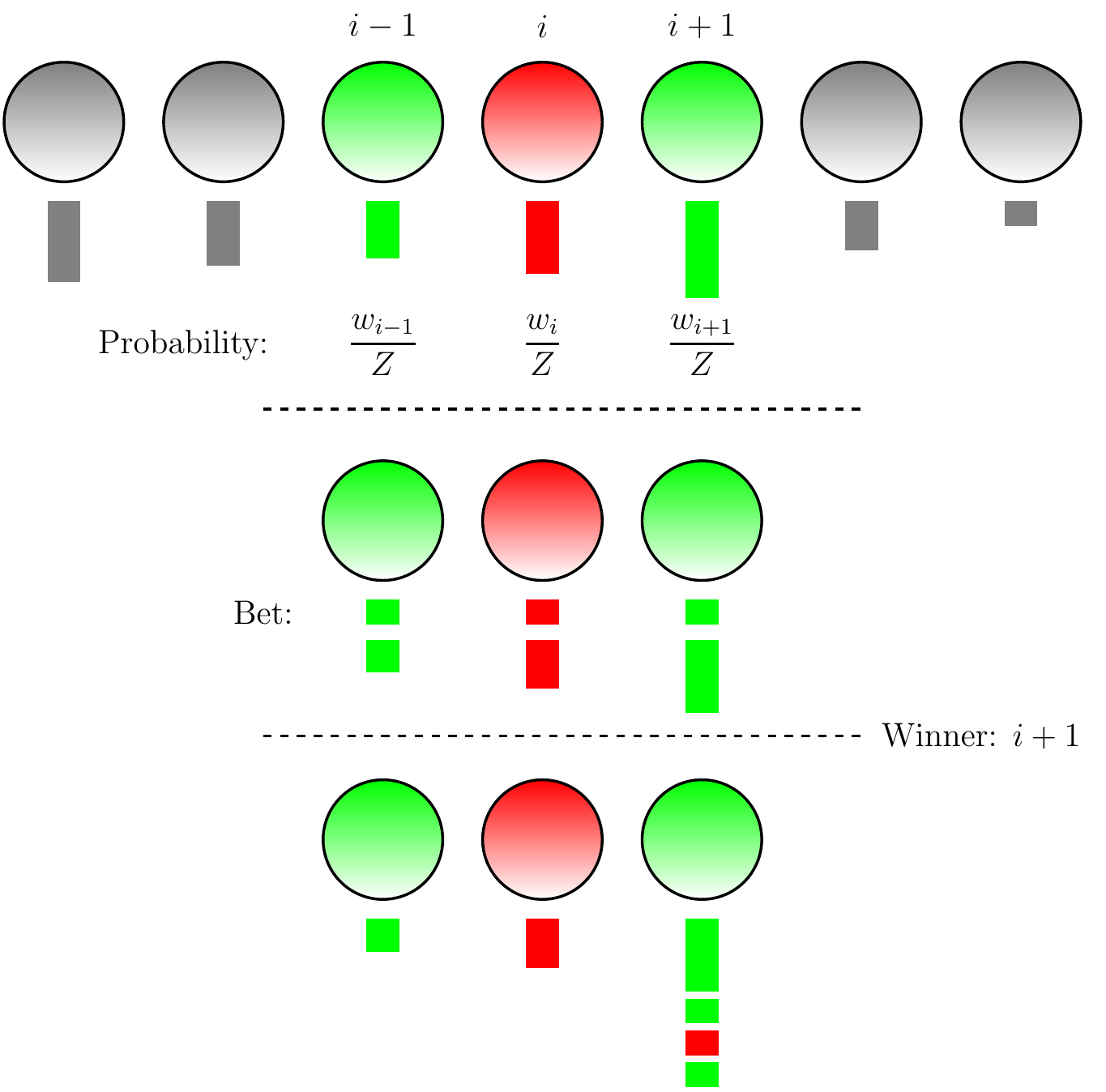}
  \includegraphics[width=8cm]{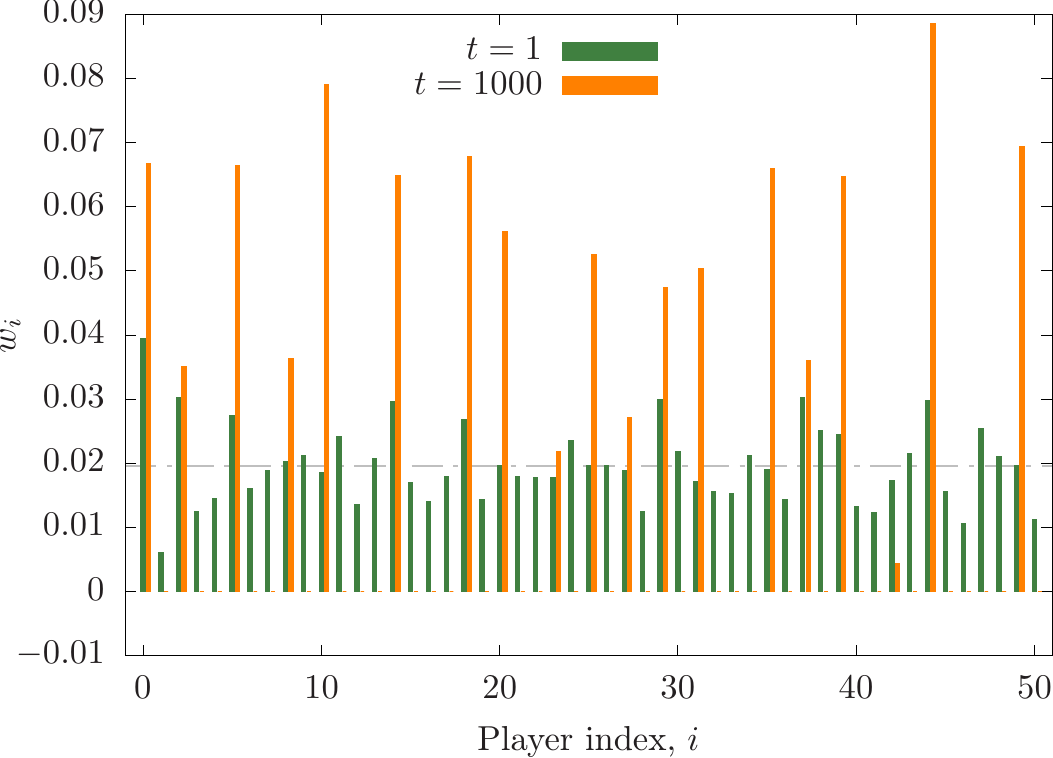}
  \caption{Top: Illustration of the power game for a one-dimensional
    graph. First stage: the selected agent is depicted in red, while
    the two neighbors appear in green, all others in gray. The bar
    under each player indicates her power. Second stage: players
    separate their bet. Third stage: a winner has been chosen (player
    $i+1$), and she receives all three bets. Bottom: two profiles for
    a system with $N=50$ and $\alpha=0.01$, for short and long times
    ($t=1$ and $t=1000$). Notice that, in the long run, high class
    agents are effectively separated by low class ones.}
  \label{fig:illust}
\end{figure}

At each round, a randomly selected agent $i$ will propose a bet to her
neighbors within the graph \footnote{We have decided to break the
  gender symmetry associated to the term {\em agent} in the female
  direction.}. The agent will bet a certain fraction of her power,
$\alpha w_i$ with $\alpha<1$, and her neighbors will be required to
call the bet. Neighbor players whose power is larger than $\alpha w_i$
are forced to do so, all others are discarded. Let ${\cal B}_i$ be the
set of {\em active} players, whose cardinal is $m_i$. A winner is
chosen with probability proportional to their power, i.e. the
probability that player $j\in {\cal B}_i$ will win is

\begin{equation}
  P_j = {w_j \over Z_i}, \qquad Z_i=\sum_{k\in {\cal B}_i} w_k.
  \label{eq:winning}
\end{equation}
After a winner is chosen, she earns an extra amount of power $\alpha
w_i(m_i-1)$, and all others reduce their power in the amount $\alpha
w_i$. Figure \ref{fig:illust} provides an illustration of the basic
{\em power game} procedure.

In order to determine a proper time scale, we will define a time-step
as a sequence of $N$ rounds of the power game. During the first time
step the homogeneous system develops random fluctuations, and these
fluctuations will grow with time. In the long run, after a time of
order $\alpha^{-1}$, most agents will be ruined, possessing negligible
power, and a few of them will concentrate nearly all the power. A
clear class division can be established in our case: agents who can
call all possible bets by their neighbors are termed {\em powerful},
and correspond to the higher class. Agents who can not call some bets
will be termed {\em powerless}, or lower class. In some situations, a
single agent may accumulate all the power for long times.

\subsection{Redistribution}

In order to diminish the drive towards inequality, we may introduce a
global redistribution mechanism, which we will call {\em taxation}.
After each time-step, all players will provide to a central authority
an amount $\tau w_i$ of their power, with a fixed $\tau<1$ that we
will call the tax rate. The total collected amount, which is equal to
$\tau$, is shared equally among them. In other words:

\beq
w_i \,\mapsto\, w_i - \tau w_i + \tau/N.
\label{eq:taxes}
\eeq
Notice that taxation reduces the power possessed by individuals whose
power exceeds the average value, $1/N$, and increases the power of the
rest. Also, notice that our proposed taxation mechanism is {\em
  non-progressive}, since all players must provide the same fraction
of their power. It is interesting to consider $\tau^{-1}$ to be the
time-scale required for the redistribution mechanism to reach an
egalitarian state, starting from any distribution.

\subsection{Measuring inequality}

The set of $\{w_i\}$ will be called a {\em profile}. We define the
{\em width} or roughness of the profile in similarity to the
definition in kinetic roughening \cite{Barabasi}:

\beq
W^2=\sum_{i=1}^N \(w_i - \frac{1}{N}\)^2,
\label{eq:width}
\eeq
where we substract $1/N$ because it is the average power possessed by
all agents. Notice that $W^2/N$ is the {\em variance} of the power.

Another useful measure of inequality is the {\em Shannon entropy} of
the profile, defined by \cite{Shannon48}. Since all powers $\{w_i\}$
are positive and add up to one, they can be regarded as a probability
distribution and we have

\beq
S=-\sum_{i=1}^N w_i \log w_i.
\label{eq:S}
\eeq
Notice that $N_\eff=\exp(S)$ can be used as an estimate for the number
of {\em active players}. In effect, if we choose an agent $k$ with
probability $w_k$, transmitting our choice will require $S$ bits,
similarly to a choice among $N_\eff$ equally likely agents.

We will also consider a sorted version of the profile, $\{\omega_i\}$,
in non-increasing order: $\omega_1\geq \omega_2 \geq \cdots \geq
\omega_N$. In other terms, $\omega_k$ is the power possessed by the
$k$-th most powerful agent. It allows us to define the fraction of
agents with power larger than $w$, $\rho(w)$, through the following
relation: $\rho(w)=k/N$ when $\omega_k=w$. Moreover, we will define
the {\em Lorenz curve} \cite{Chatterjee15}, ${\cal L}(k/N)$ is the
total power possessed by the poorest $k$ players:

\beq
    {\cal L}(k/N)=\sum_{i=1}^k \omega_{N-i+1}.
\label{eq:lorenz}
\eeq
The Lorenz curve allows to define the {\em Gini coefficient}
\cite{Chatterjee15} of the distribution as the area between the Lorenz
curve and the perfect equality value, ${\cal L}(x)=x$.

\beq
G\equiv  \int_0^1 dx  \( x-{\cal L}(x)\).
\label{eq:gini}
\eeq
It can be shown to correspond to the expected value of the differences
of power of two different players, multiplied by $N$,

\beq
G=N \sum_{i,j=1}^N |w_i-w_j|.
\label{eq:gini2}
\eeq

In order to gather some intuition, let us consider a case where $N_u$
upper class agents concentrate all the power among themselves,
equally. The width of the distribution can be found applying
Eq. \eqref{eq:width},

\beq
W^2={1\over N_u}-{1\over N}.
\eeq
The entropy of the distribution can be found to be $S=\log(N_u)$,
expressing the amount of information required to transmit the identity
of an agent selected with a probability equal to her power. On the
other hand, the Gini index can be found using Eq. \eqref{eq:gini}, and
it is

\beq
G=1-{N_u\over N},
\eeq
i.e. equal to the fraction of lower-class individuals.


\section{Two Players Game}
\label{sec:twoplayers}

Let us consider the case with two players, with powers $w_1$ and
$w_2$. With some abuse of notation, let us call $w_i(t)$ the {\em
  expected} value for the power of the first player at time $t$. The
expected amount of the bet is $\alpha/2$ ($\alpha w_i(t)$ when placed
by player $i$). If she wins, she will gain an amount of power $\approx
\alpha$, but only with probability $w_i(t)$. Thus, the expected value
for $w_i(t+1)$ is given by

\beq
w_i(t+1) = w_i(t) + \alpha \(w_i(t) - \frac{1}{2} \).
\label{eq:twop_mf}
\eeq

This equation can be iterated, and we obtain

\beq
w_i(t) = \frac{1}{2} + \(w_i(0)-\frac{1}{2}\) (1+\alpha)^t.
\eeq
In other words, the expected value of the power will increase 
(decrease) exponentially when the initial value is above (below) 1/2,
$\exp(\pm \gamma t)$, with $\gamma=\log(1+\alpha)$.

This approach does not inform us about the {\em fluctuations}. Thus, a
more detailed calculation is needed. Table \ref{table:2p} shows the
payoff matrix for a single power game with two players, depending on
who is the gambler and who is the winner.

\begin{table}[htbp]
\centering
\begin{tabular}{|c|c||c|c|}
  \hline
 Gambler & Winner & Probability & Payoff\\
 \hline \hline
 $1$ & $1$ & $\frac{1}{2} w_1$ & $+\alpha w_1$\\
\hline
 $1$ & $2$ & $\frac{1}{2} (1-w_1)$ & $-\alpha w_1$\\
 \hline
 $2$ & $1$ & $\frac{1}{2} w_1$ & $+\alpha (1-w_1)$\\
\hline
 $2$ & $2$ & $\frac{1}{2} (1-w_1)$ & $-\alpha (1-w_1)$\\
\hline
\end{tabular}
\caption{Payoff matrix for a single round of the power game between
  two players.}
\label{table:2p}
\end{table}

The expected gain of player 1 is given by the following expression:

\begin{eqnarray}
\textrm{E}(\textrm{Gain}) &=& \sum \textrm{Prob.} \times \textrm{Payoff}
= \alpha \(w_1-\frac{1}{2}\).
\label{eq:2paver}
\end{eqnarray}
where the sum is implicitly extended over all possible outcomes.

In order to obtain the variance of the gain of player 1, we make use
of Table \ref{table:2pvar}, where we describe the expected squared
deviations with respect to the expected payoff. Thus, the variance of
the payoff of player 1 is given by

\begin{table}[htbp]
\centering
\begin{tabular}{|c|c||c|c|c|}
  \hline
 Gambler & Winner & Probability & $($Payoff$-$E$($Payoff$))^2$\\
 \hline \hline
 $1$ & $1$ & $\frac{1}{2} w_1$ & $\alpha^2/4$ \\
\hline
 $1$ & $2$ & $\frac{1}{2} (1-w_1)$ & $\alpha^2(4w_1-1)^2/4$\\
 \hline
 $2$ & $1$ & $\frac{1}{2} w_1$ & $\alpha^2(4w_1-3)^2/4$\\
\hline
 $2$ & $2$ & $\frac{1}{2} (1-w_1)$ & $\alpha^2/4$\\
\hline
\end{tabular}
\caption{Fluctuations in the payoff of player 1.}
\label{table:2pvar}
\end{table}

\begin{eqnarray}
  \textrm{V}(\textrm{Payoff}) =
  \sum \textrm{Prob.} \times
  \(\textrm{Payoff}-\textrm{E}(\textrm{Payoff}) \)^2 = \frac{\alpha^2}{4}
\label{eq:2pdev}
\end{eqnarray}

\begin{figure*}
  \hbox to 16cm{
  \includegraphics[width=6.5cm]{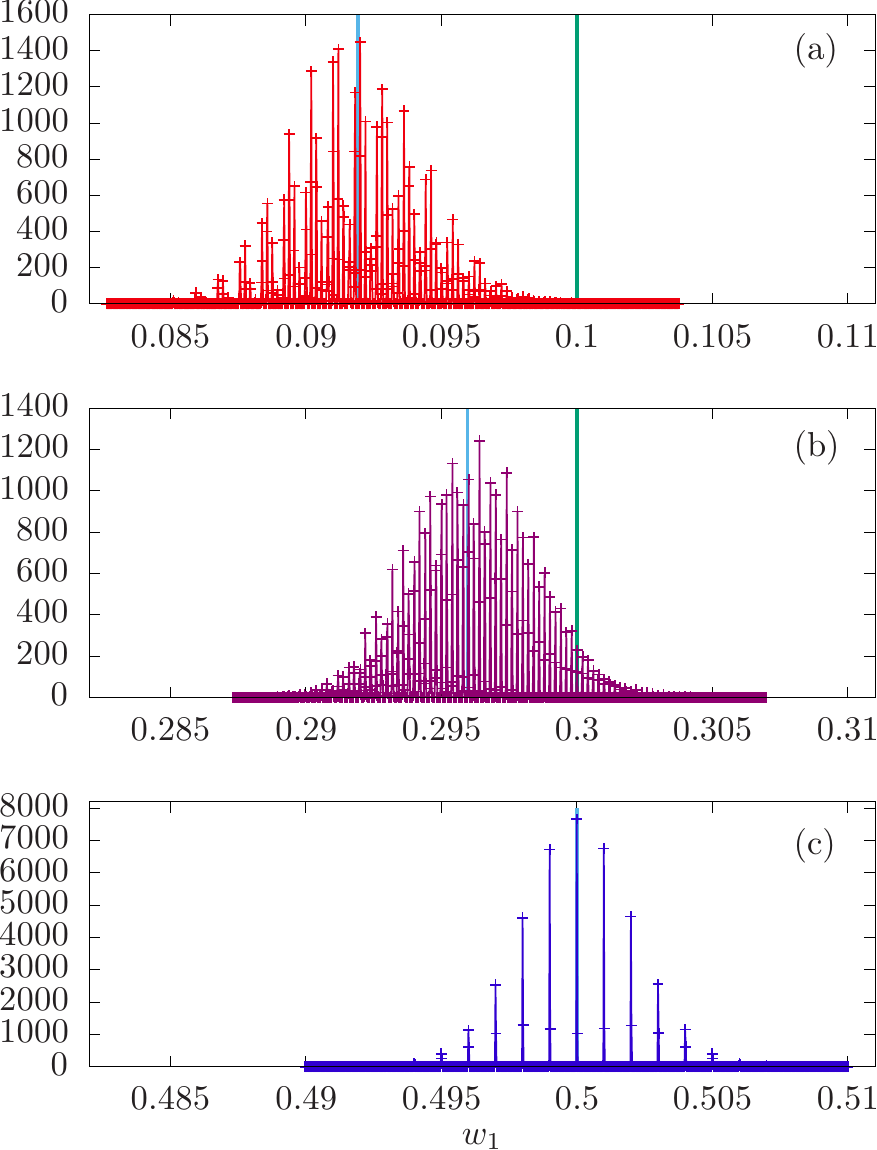}\hfill
  \includegraphics[width=7cm]{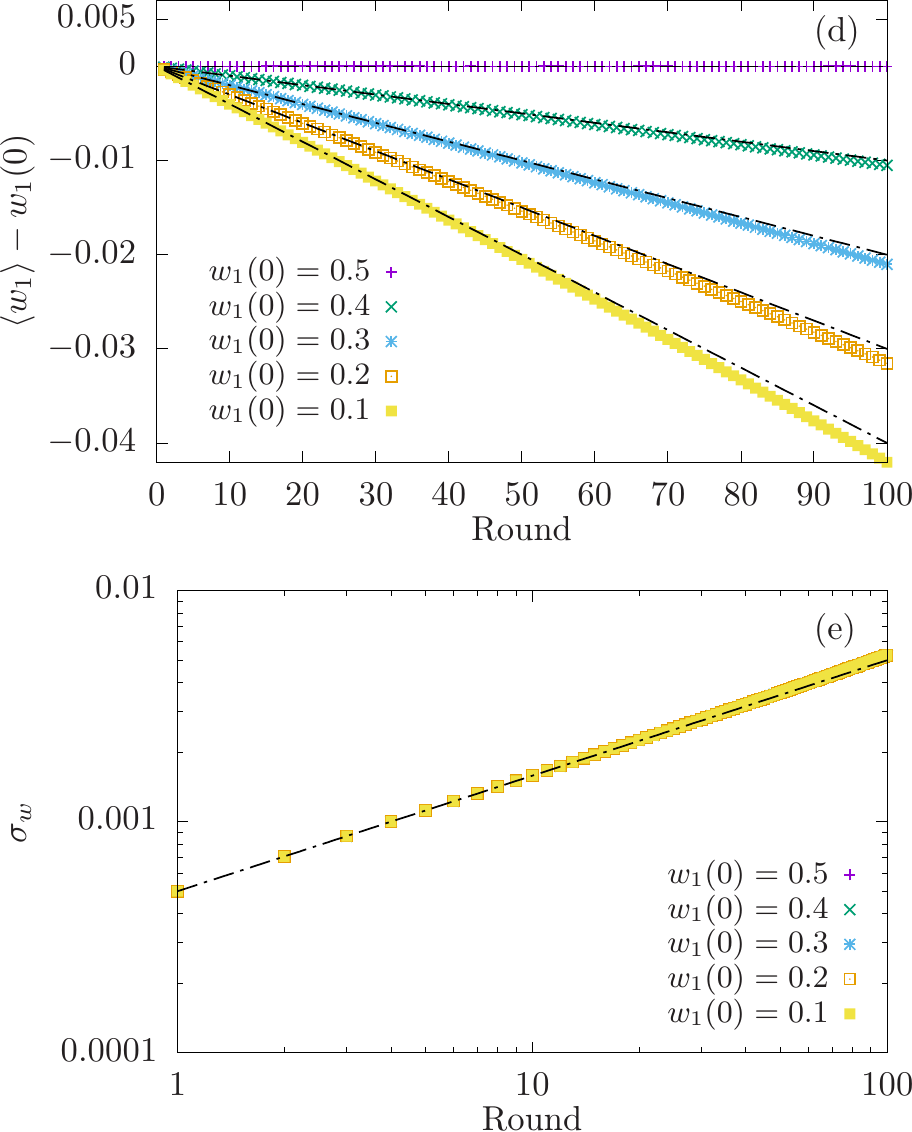}}
  \caption{Left: Two-player histogram of $w_1$ using $\alpha=10^{-3}$
    for different initial values $w_1(0)=0.1$ (a), $0.3$ (b) and $0.5$
    (c) and 20 games, using $10^6$ realizations for each case. (d)
    Average power of player 1 for the different initial conditions,
    $w_1(0)=0.1$ up to $0.5$ as a function of the round order. (e)
    Power deviation for the same cases. Notice the collapse of all
    curves, as predicted in Eq. \eqref{eq:2pdev}. Dashed lines are
    theoretical predictions, Eqs. \eqref{eq:2paver} and (the square
    root of) \eqref{eq:2pdev}.}
  \label{fig:juego1}
\end{figure*}

Both predictions, Eq. \eqref{eq:2paver} and \eqref{eq:2pdev}, are
checked against numerical simulations in Fig. \ref{fig:juego1} (d) and
(e), for a small number of rounds (100) and a low value of
$\alpha=10^{-3}$. The fit is remarkably accurate.

Interestingly, the collapse of all deviation curves shown in
Fig. \ref{fig:juego1} (e) and predicted in Eq. \eqref{eq:2pdev}
depends on the precise rules employed. Appendix \ref{sec:appendix}
discusses a slightly different set of rules for which this collapse
does not take place.
\vspace{2mm}

Notice that, in the long run, the system can be in two different
absorbing states: one of the players will possess all the power while
the other will be ruined. The probability of ever visiting the other
state becomes negligible as the game length increases. This implies a
breaking of ergodicity \cite{LZ1,LZ2,Berman17,Stojkovski19}: the
stationary state of the system depends on its (early) history, and can
not be properly called an equilibrium state.


\section{One-Dimensional Power Game}
\label{sec:1d}

The power game, as described in Sec. \ref{sec:powergame}, is defined
on an arbitrary graph. We will consider in this section the
application to a number $N$ of players displayed along a ring. For all
cases, we have run $100$ simulations and averaged the results,
checking for convergence.

\subsection{Without Taxes}

\begin{figure}
  \includegraphics[width=8cm]{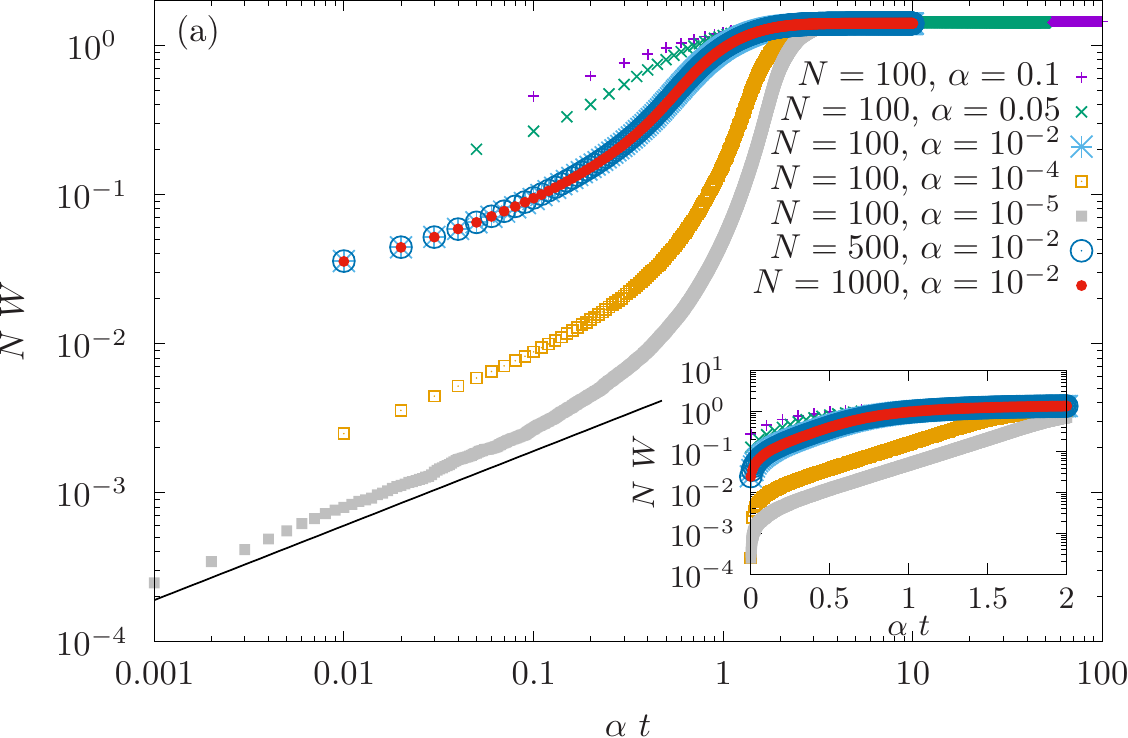}
  \vskip 5mm
  \includegraphics[width=8cm]{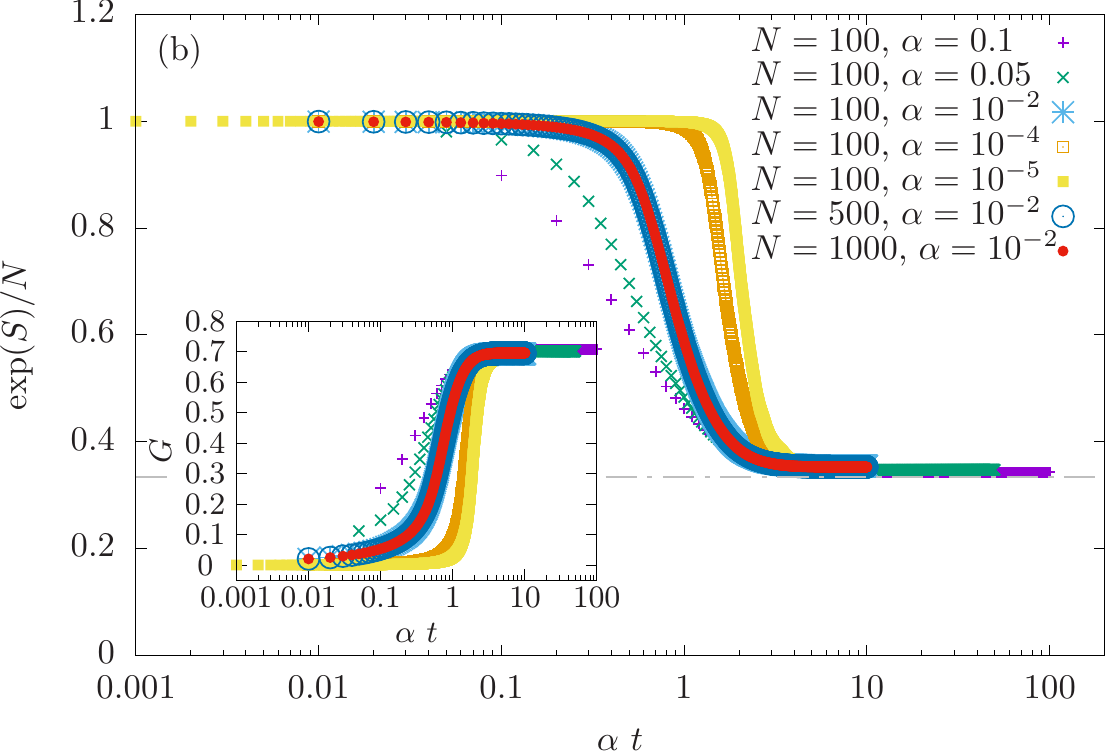}
  \caption{(a) Average width of the power distribution of the 1D
    power game as a function of time, for different values of $N$ and
    $\alpha$. In all cases, the tax level $\tau=0$. The time axis is
    rescaled as $\alpha t$, while the width axis is rescaled as
    $NW$, in order to highlight the collapse at
    saturation. (b) entropy decrease as a function of time, with
    the time-axis rescaled as $\alpha t$, and the entropy rescaled as
    $\exp(S)/N$, because this magnitude corresponds to the fraction of
    high class agents. Notice that this fraction comes close to $1/3$
    in the long run.}
  \label{fig:linWS}
\end{figure}

The average width of the power distribution $W$, given by
Eq. \eqref{eq:width}, is shown as a function of time for different
numbers of players and values of $\alpha$ in
Fig. \ref{fig:linWS} (a). In all cases, the tax level was set to
zero. The time axis is rescaled, as $\alpha t$, while the width axis
is also rescaled, as $N W$. Notice that all curves start at zero $W$,
reaching a saturation level for long times. The rescaling allows all
curves to have the same saturation, independently of $\alpha$ and $N$,
and a very similar saturation time. For short times, $W$ grows as a
power law, $W\sim t^{1/2}$, in agreement with Family-Vicsek scaling
for kinetic roughening \cite{Barabasi} in the case of {\em random
  deposition}. Yet, before saturation the width increases much
faster. The inset shows that this second stage of growth is, indeed,
exponential, which does not correspond to Family-Vicsek scaling.

The evolution of the Shannon entropy $S$, Eq. \eqref{eq:S}, is shown
in Fig. \ref{fig:linWS} (b). The time-axis is again rescaled as
$\alpha t$, while the entropy is shown as $\exp(S)/N$, in order to
make all curves collapse, for different values of $N$ and $\alpha$. As
it was argued, $\exp(S)/N$ denotes the fraction of upper class agents,
which converges to $\approx 1/3$ for long times.

The inset of Fig. \ref{fig:linWS} (b) shows the time
evolution of the Gini index, with time again scaled as $\alpha t$. We
notice that it evolves from $G=0$ to $G\sim 0.7$. In the class
division picture, this value corresponds to the fraction of lower
class agents, which should be close to $2/3$.

Thus, our observations are compatible with the conjecture that, for
long times, the system is divided into an upper class, containing
$\approx 1/3$ of the agents, roughly similar in power, and a lower
class, with the rest of the agents. The lower class agents possess
negligible power, and are unable to call the bets placed by the higher
class agents. Thus, the origin of the 1/3 fraction should be
understood in the following way: a stationary state is reached if each
high class agent is surrounded by two lower class ones which are not
neighbors of any other high class agent. A typical distribution with
this property can be built by alternating one high class and two lower
class agents, thus giving rise to the observed division.

Moreover, two high class agents can never be neighbors in the
stationary state. That situation is unstable: in the long run, one of
them will always grab the power of the other. So, when we start from a
homogeneous state, the mean value of the separation between high class
agents will be two in the long run. Yet, stationary states are not
perfect crystals, and this separation is sometimes higher or
lower. Even more: if we are allowed to start from an inhomogeneous
situation, we can tune this separation to higher values. For example,
an initial state with all power in the hands of one of the agents will
remain stationary. Thus, we insist that our model is strongly
non-ergodic.
\vspace{3mm}

\begin{figure}
  \includegraphics[width=8cm]{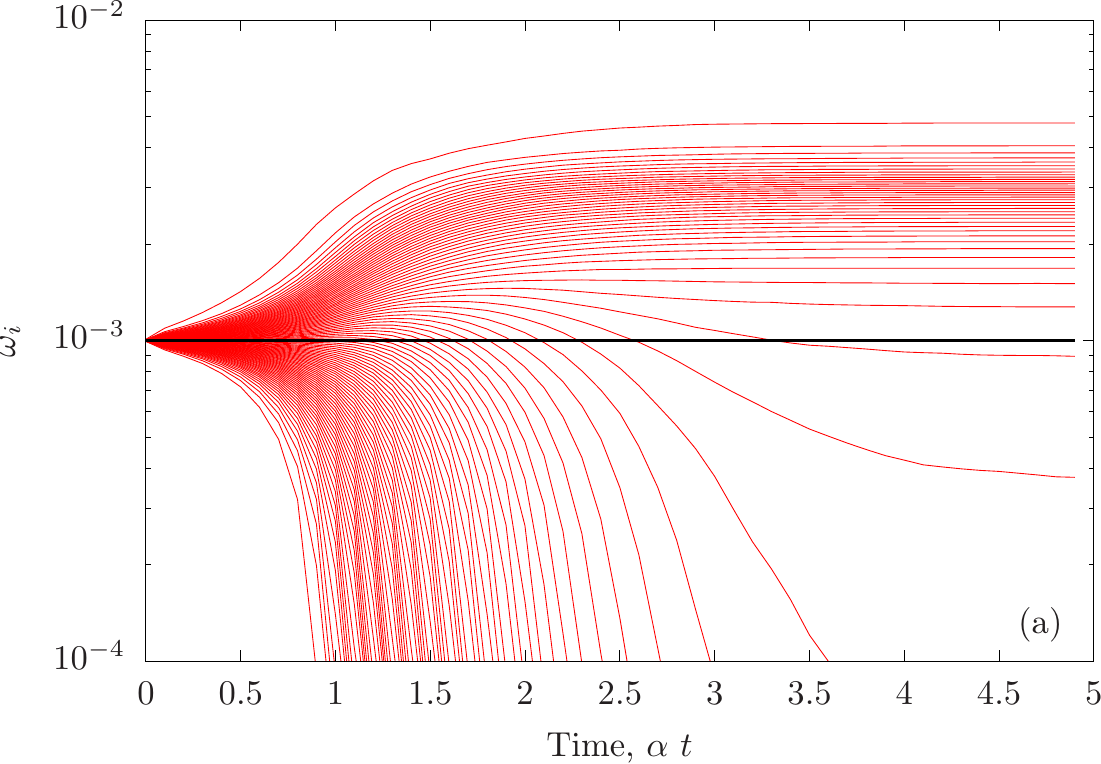}
  \vskip 5mm
  \includegraphics[width=8cm]{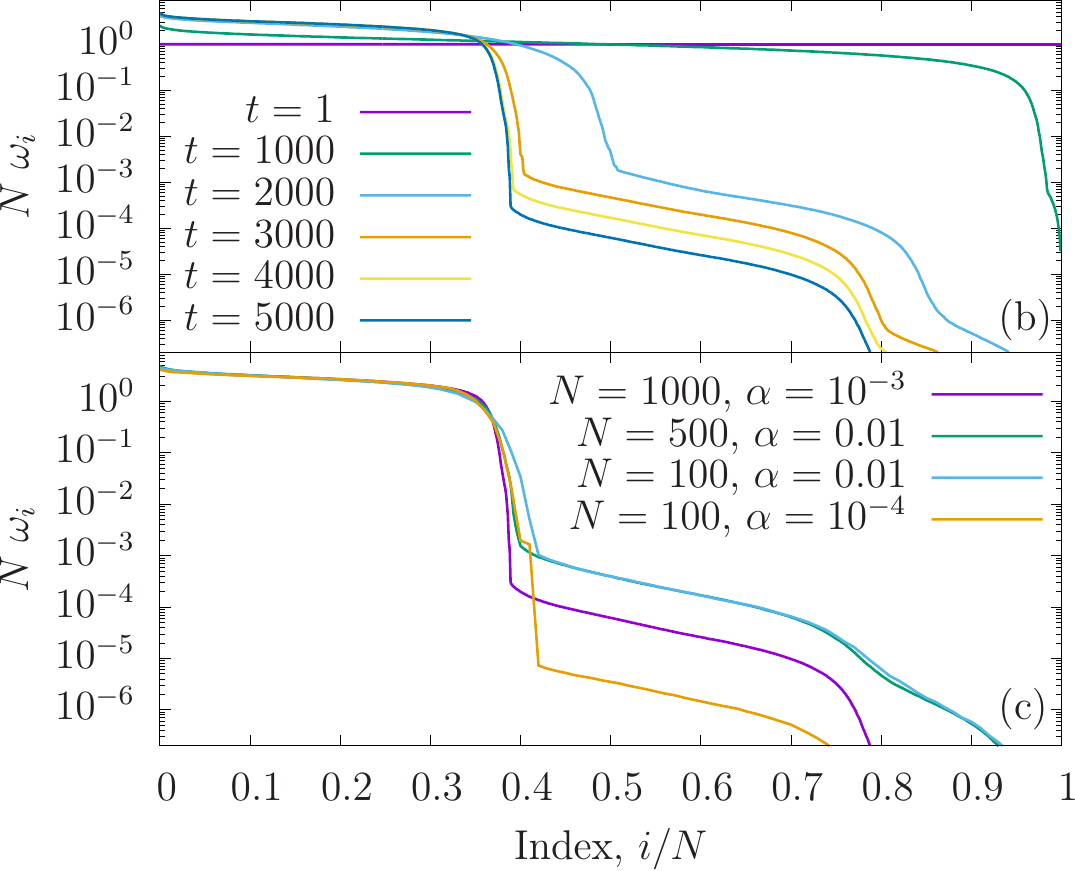}
  \caption{(a) Time evolution of selected values of the sorted
    profile, $\{\omega_k\}$ with $k=10m+1$, using $N=10^3$ and
    $\alpha=10^{-3}$. The time axis is rescaled as $\alpha t$. Notice
    the initial exponential divergence, followed by saturation around
    $\alpha t \sim 1$. (b) and (c) Full sorted profile as a function of
    the index $k$, with suitably rescaled axes, for different values
    of time (b) and different system sizes and values of $\alpha$ in
    the stationary state (c).}
  \label{fig:linomega}
\end{figure}

We can obtain further insight about the power distribution by
considering the time evolution of the {\em sorted profile}:
$\{\omega_k\}$, as shown in Fig. \ref{fig:linomega} (a)
for $N=10^3$ and $\alpha=10^{-3}$. Not all values of $\omega_k$ are
shown, only one out of every 10: $k=10m+1$. In an initial short time
regime, the values diverge exponentially from the initial one,
$1/N$. But for $\alpha t\sim 1$, the maximal probability $\omega_1$
saturates. Even before, many of the lesser agents have started an
extremely fast decay.

In Fig. \ref{fig:linomega} (b) and (c) we show the full sorted
profile, $\{\omega_k\}$. Fig. \ref{fig:linomega} (b) shows these
sorted profiles for different times for the previous system, $N=10^3$
and $\alpha=10^{-3}$. Notice that, for very short times, $N\omega_k
\sim 1$, while it grows more and more unequal for larger times. The
stationary curve has been obtained with a good approximation for
$t=5000$, the last curve. For $t=1$ the sorted profile is totally
flat. We see that two classes develop from the beginning, separated by
a large step and with lower internal variance. As time evolves, the
top part of the high class barely change, but a few high class agents
fall into the lower class, as we see from the leftward shift of the
class boundary with time. Fig. \ref{fig:linomega} (c) shows the
stationary sorted profiles reached for different values of $N$ and
$\alpha$.

The collapse of the top part of the profiles show that the parameter
$\alpha$ determines the time-scale, while $N$ determines the length
scale. The continuum limit is obtained for $\alpha\to 0$ and
$N\to\infty$.

\subsection{With Taxes}

Let us determine the effects of redistribution, through the use of
{\em taxes}, as shown in Eq. \eqref{eq:taxes}. As we will see, a
transition takes place for $\tau\sim \alpha$, from an unequal
distribution to a homogeneous one, with some unexpected effects in the
intermediate region.

\begin{figure}
  \includegraphics[width=8cm]{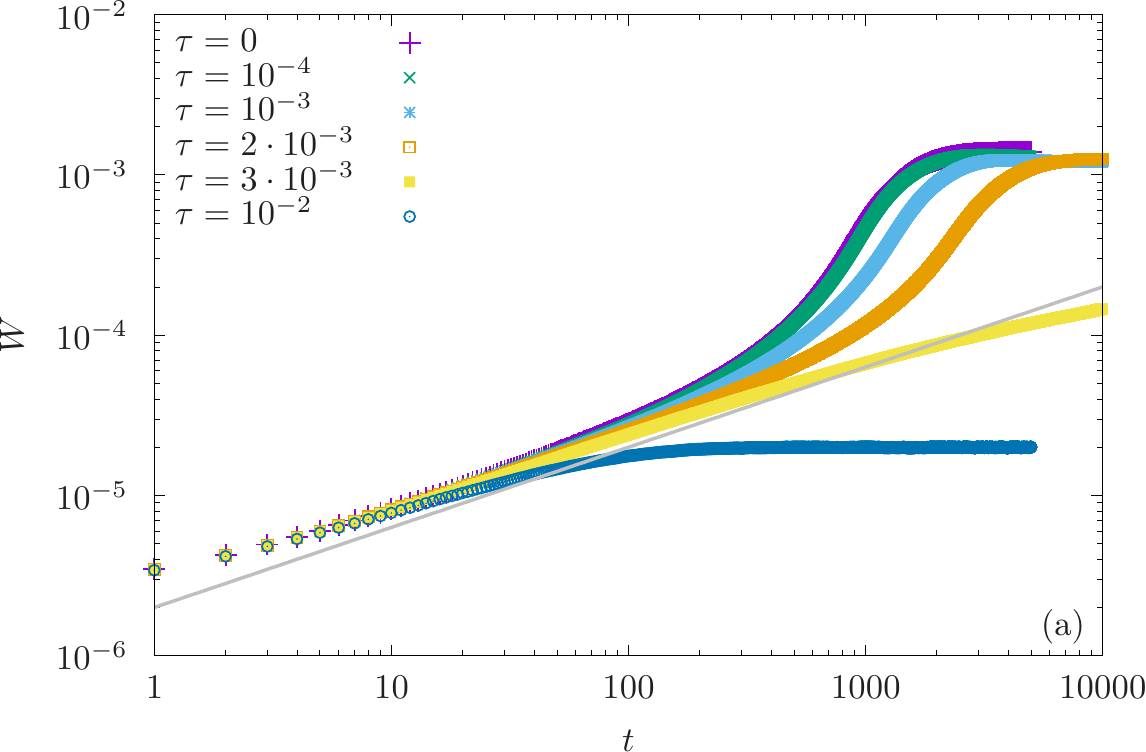}
  \vskip 5mm
  \includegraphics[width=8cm]{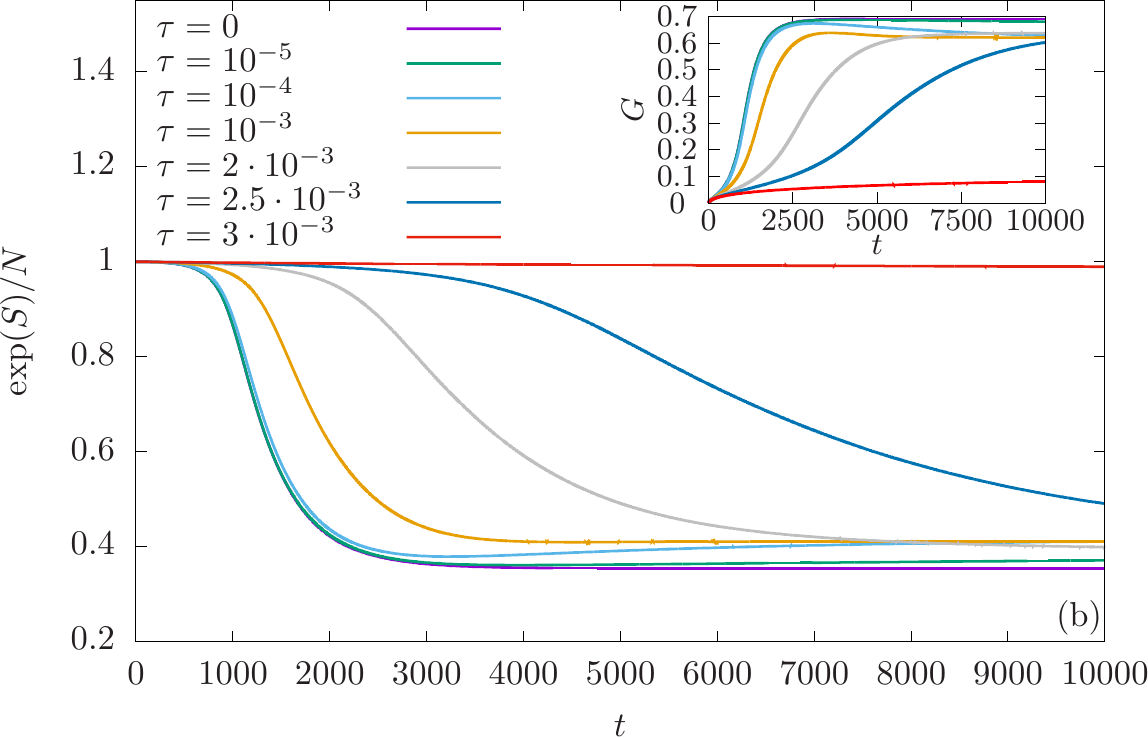}
  \includegraphics[width=8cm]{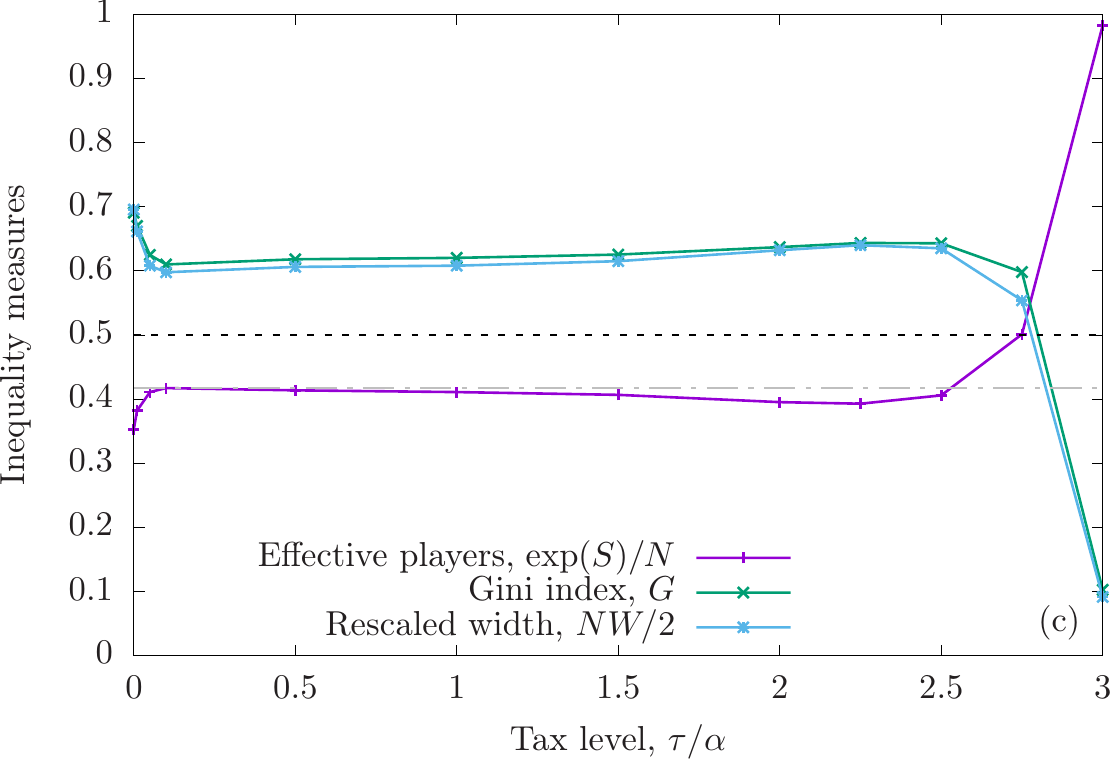}
  \caption{(a) Average width of the 1D power game with taxes, using
    $N=10^3$ and $\alpha=10^{-3}$. The gray line stands for a
    $t^{1/2}$ growth. (b) Time evolution of the entropy and Gini
    index of the distributions. (c) Stationary values of the
    number of effective players, Gini index and roughness for the same
    systems as a function of the tax level $\tau/\alpha$.}
  \label{fig:taxWS}
\end{figure}

Fig. \ref{fig:taxWS} shows the width, entropy and Gini for a system
with $N=10^3$, $\alpha=10^{-3}$ and different taxation levels. In
Fig. \ref{fig:taxWS} (a) we can see that the width, $W$, grows for
short times as $W\sim t^{1/2}$ in all cases. For low taxation levels,
$\tau<3\alpha$, it still presents the same features of the no-taxation
case, with the accelerated growth of the inequality near saturation,
and the saturation levels reached are similar. But above that
threshold value, the behavior changes drastically, with the saturation
level falling sharply after a small increase in the taxation level.

Fig. \ref{fig:taxWS} (b) provides further information about this
transition. We see the time evolution of the effective number of
players, or size of the high class, estimated from the entropy,
Eq. \eqref{eq:S}. Indeed, for $\tau< 3\alpha$ this number decays
steadily down to a value of order $N/3$. Yet, for $\tau \geq 3\alpha$,
the effective number of players does not seem to decay, even in the
long run. Thus, we conjecture that the system is not divided into
classes any more. The Gini index, depicted in the inset of
Fig. \ref{fig:taxWS} (b), shows the same behavior: it grows steadily
to a value close to $0.7$ for $\tau < 3\alpha$, and remains close to
zero for $\tau \geq 3\alpha$.

Yet, a careful analysis of Fig. \ref{fig:taxWS} (a) and (b) holds a
surprise more. Indeed, the long term $N_\eff$ is not a monotonously
increasing function of $\tau$: for $\tau=\alpha$ it takes a larger
value than for $\tau=2\alpha$, although just marginally so. This
effect is checked carefully in Fig. \ref{fig:taxWS} (c), where the
stationary (long term) values of the three previous observables (width
$W$, entropy $S$ and Gini $G$) are plotted as a function of the
taxation level, $\tau/\alpha$. Interestingly, we observe in all three
the same trend: a low taxation brings about a mild decrease in
inequality. Yet, increasing the taxation level from $\tau=\alpha/2$ to
$\tau=2\alpha$ increases the inequality a tiny amount. If we increase
the taxation level further, inequality disappears sharply. The reason
seems to be that mild levels of taxation may have a counterintuitive
effect on inequality: by providing enough power to the worse-off
agents, the high class agents are not effectively isolated and can
still compete among themselves. Yet, we must stress that this slight
uptake in the inequality rate as taxes grow is not a robust trait of
the model, as we have not been able to reproduce it in higher
dimensional lattices (see Sec. \ref{sec:2d}).

\vspace{3mm}

\begin{figure}
  \includegraphics[width=8cm]{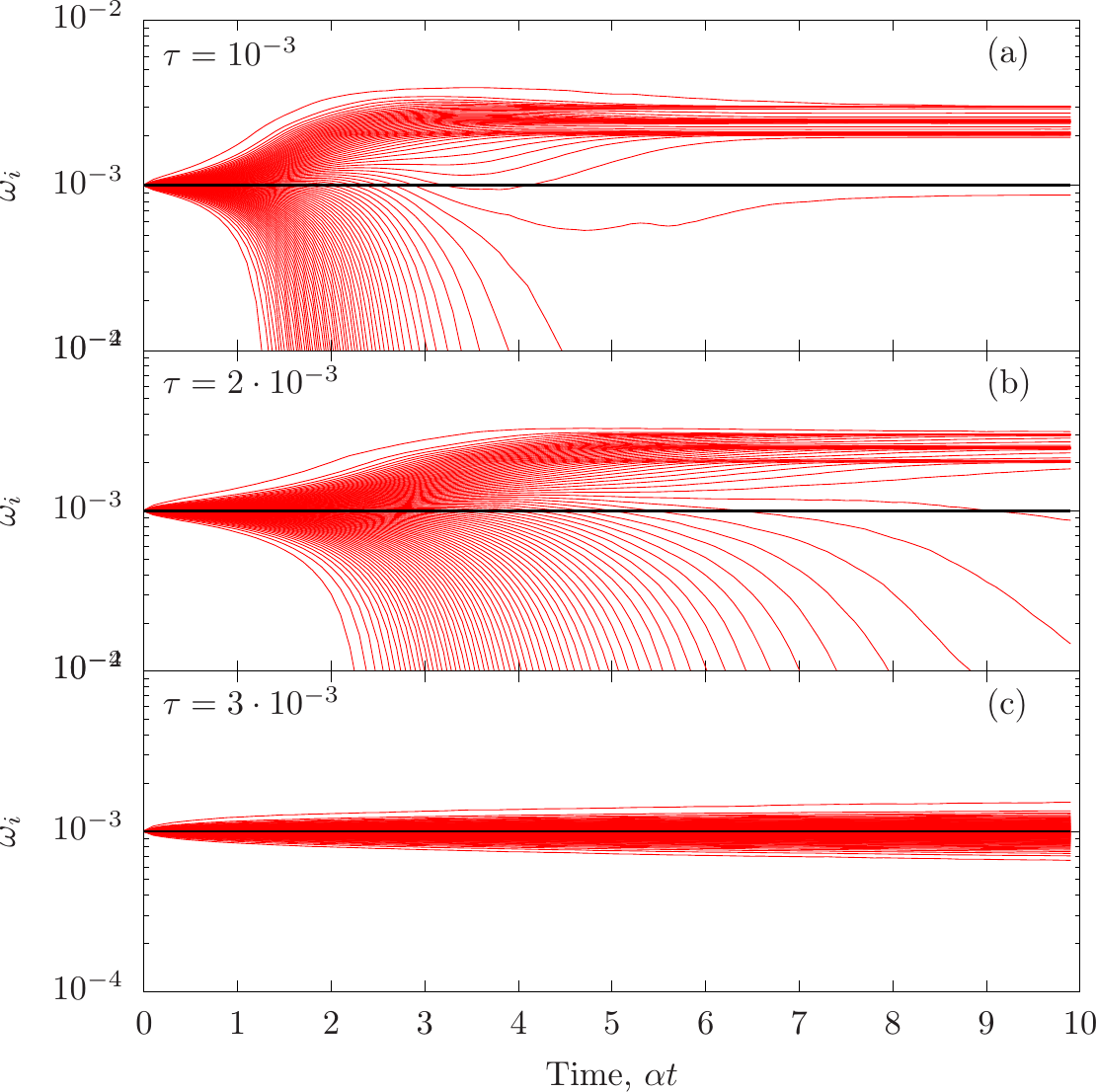}
  \vskip 5mm
  \includegraphics[width=8cm]{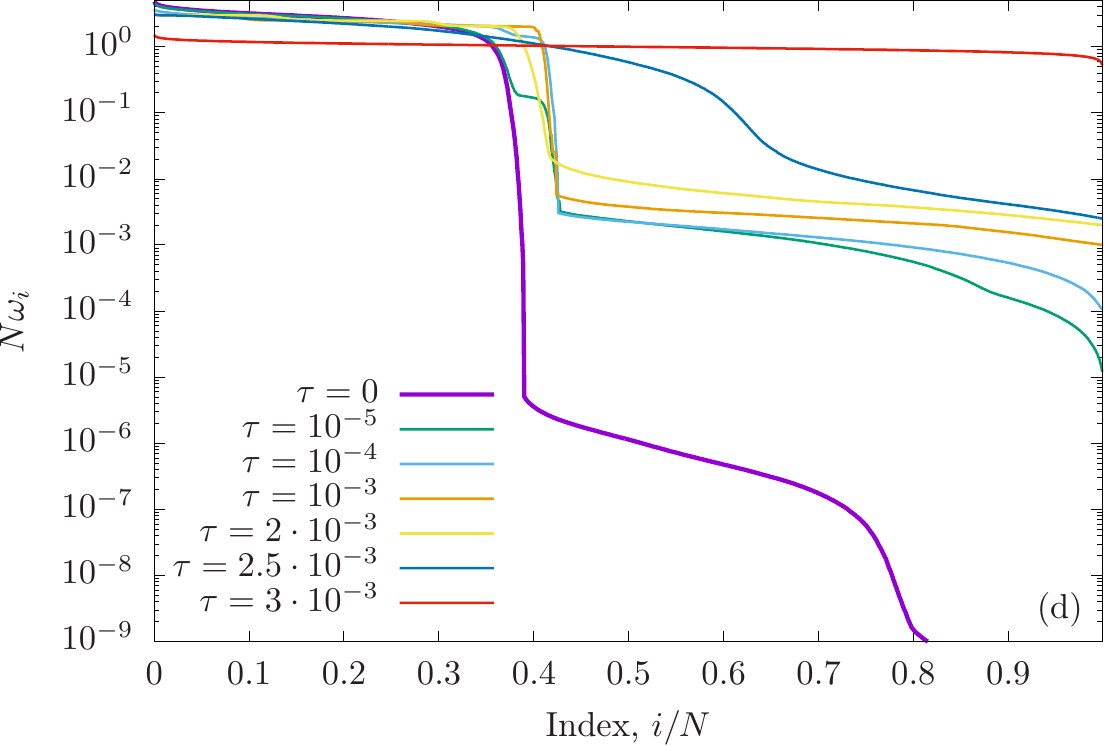}
  \caption{(a) to (c) Time evolution of the sorted power values,
    $\omega_k$ for selected values of $k$, with $N=10^3$ and
    $\alpha=10^{-3}$ and different taxation levels. The time axis is
    rescaled as $\alpha t$. (d) Stationary values of the sorted
    profiles for different taxation levels.}
  \label{fig:taxomega}
\end{figure}

Fig. \ref{fig:taxomega} (a), (b) and (c) show the time evolution of
selected values of $\omega_k$, for $k=10m+1$. Notice that for
$\tau=\alpha$, Fig. \ref{fig:taxomega} (a), the evolution is very
similar to the case without taxes. For $\tau=2\alpha$,
Fig. \ref{fig:taxomega} (b), the time axis seems to be stretched, but
inequality builds up anyway. For $\tau=3\alpha$,
Fig. \ref{fig:taxomega} (c), on the other hand, the system is
basically egalitarian, with a negliglible spread.

Fig. \ref{fig:taxomega} (d) shows the stationary
sorted profiles, $\{\omega_k\}$ for different taxation levels. For
$\tau=3\alpha$ the profile is basically flat, but we can see that the
situation can be involved for other taxation levels. 
\vspace{3mm}

\begin{figure}
  \includegraphics[width=8cm]{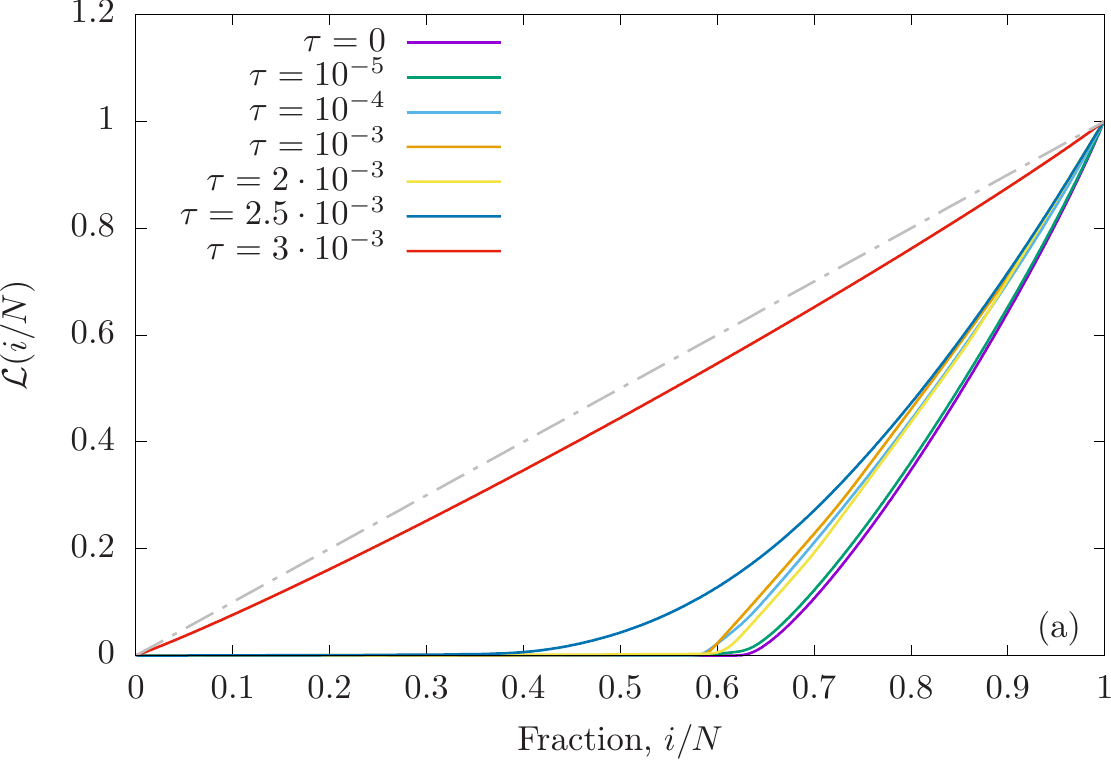}
  \vskip 5mm
  \includegraphics[width=8cm]{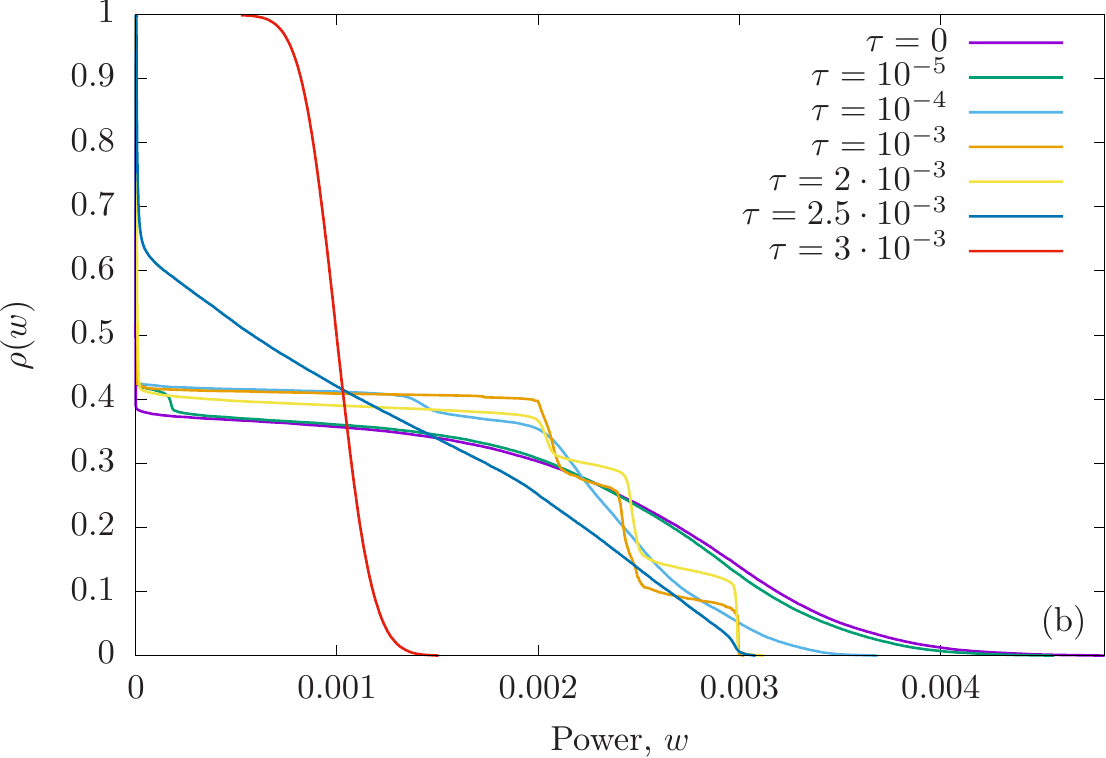}
  \caption{(a) Lorenz curves for the 1D power game with different
    levels of taxation, using $N=10^3$ and $\alpha=10^{-3}$. (b)
    Probability of finding an agent with power larger than $w$, as a
    function of $w$, for the same systems.}
  \label{fig:lorenz}
\end{figure}

Within the economics literature, two curves present special relevance:
${\cal L}(i/N)$, the Lorenz curve, defined in Eq. \eqref{eq:lorenz}
and the wealth function, $\rho(w)$, defined as the probability of
finding an agent with power larger than $w$. The interest of these
functions, which are not usual in the physics literature, is
highlighted in Fig. \ref{fig:lorenz}. In Fig. \ref{fig:lorenz} (a) we
see the Lorenz curve for the aforementioned cases. Notice that the
diagonal line corresponds to the case of perfect equality, while the
area between the curve and the diagonal line is the Gini index. We
notice that for very low taxation level the Lorenz curve is barely
different from the no-taxation one, while for larger values the system
changes drastically in a short span of taxation levels.

The $\rho(w)$ function is shown in Fig. \ref{fig:lorenz} (b) for the
same systems. We also see the strong difference achieved through a
very small change in the taxation level for $\tau\sim\alpha$. For
$\tau=3\alpha$ the wealth distribution $\rho(w)$ has a sharp decline
around $w=1/N$, showing a basically egalitarian system. Yet, for
$\tau\sim \alpha$ the curve becomes more involved, with several steps,
reflecting the complex behavior of the stationary distribution for low
taxation levels.

\vspace{2mm}

It is relevant to ask whether these results are robust under the
presence of noise. We have performed numerical experiments with {\em
  noisy tax rates}, i.e. the tax rate for each agent becomes a random
variable, uncorrelated both in space and in time. Instead of a fixed
value $\tau$, the tax rate for each player at each time-step becomes
$\tau_i=\tau(1+D\eta)$, where $D\geq 0$ is a parameter and $\eta$ is a
uniform variate in $[-1,1]$. The results (not shown) are virtually
identical even for values of $D\sim 1/2$. Indeed, the average value of
the taxation level seems to be the only relevant variable.


\section{The Power Game beyond 1D}
\label{sec:2d}

It is relevant to ask about the statistical behavior of the power
distribution when the agents interact through a more complex
network. In this section we will consider 1D and 2D networks with
interactions beyond nearest neighbors. Concretely, we will consider 1D
rings with $N$ agents, and interaction range $r=1$, 2 or 3, and
periodic 2D square lattices with $L_x\times L_y=N$ agents with
interaction range $r=1$ (nearest neighbors) and $r=2$ (nearest and
next-nearest neighbors).

\begin{figure}
  \includegraphics[width=8cm]{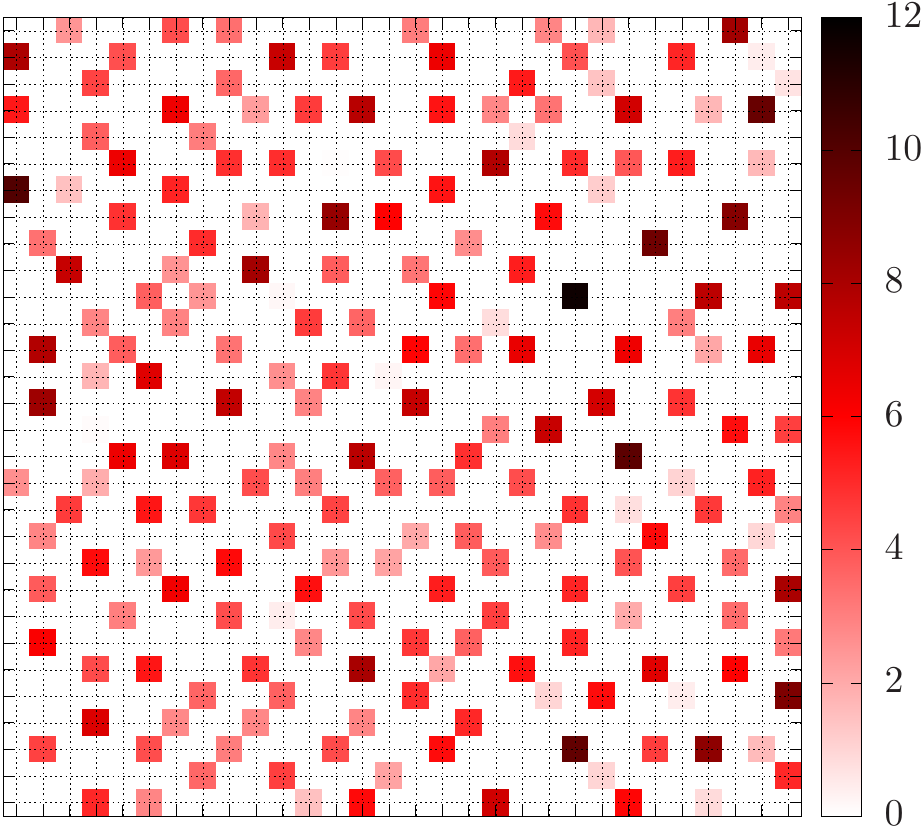}
  \caption{Power profile in the steady state for a power game played
    on a $30\times 30$ square lattice with periodic boundaries. The
    color index is $Nw$, where $w$ is the power of each agent,
    implying that $Nw=1$ corresponds to the initial equitative
    level. Notice that no high-class agent is neighbor to another one,
    but otherwise the arrangement is disordered.}
  \label{fig:illust_2d}
\end{figure}

Fig. \ref{fig:illust_2d} shows the long-term stationary state of a
power game played on a $30\times 30$ lattice with periodic boundary
conditions and range $r=1$, with $\alpha=0.01$ and no taxes. The
colorbox is scaled as $Nw$, where $w$ is the power of each agent, so
that $Nw=1$ correspond to the initial equitative distribution. Players
are still divided into two classes, and two high class players can not
be together along a horizontal or vertical line in the stationary
state. Yet, the configuration does not appear to present any type of
long-range order. Remarkably, despite the apparent lack of a local
order parameter, the number of high class players is one fifth of the
total population, with a rather good precision.

\begin{figure}
  \includegraphics[width=8cm]{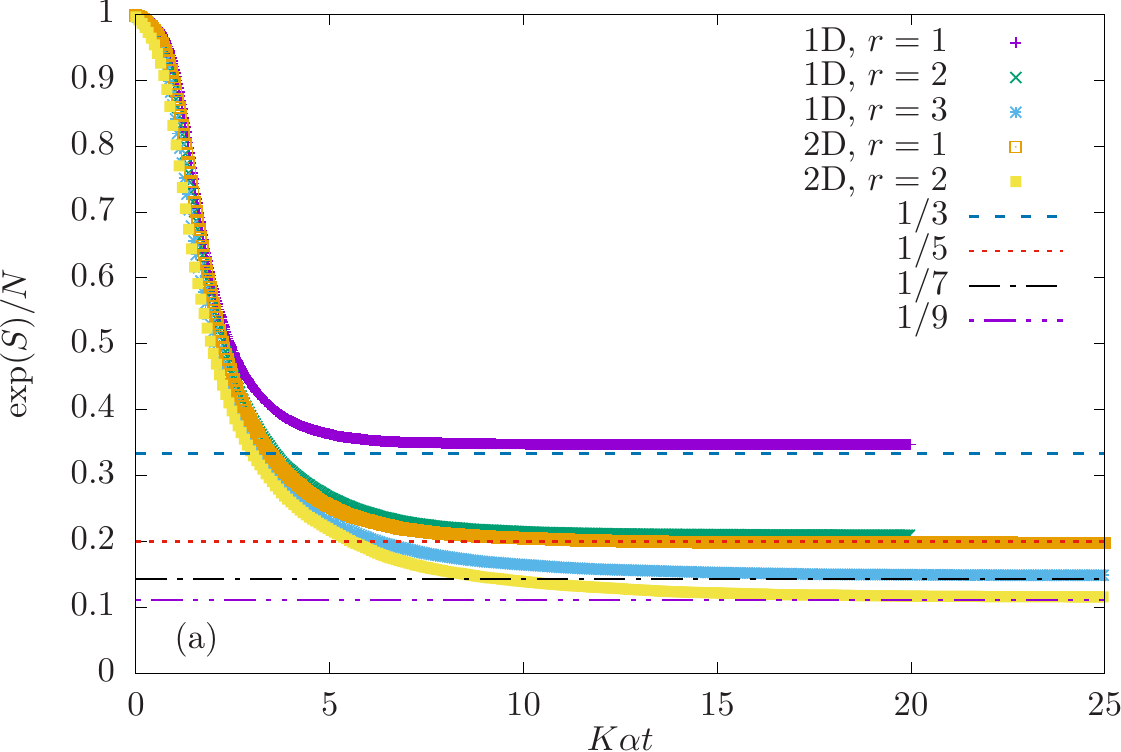}
  \includegraphics[width=8cm]{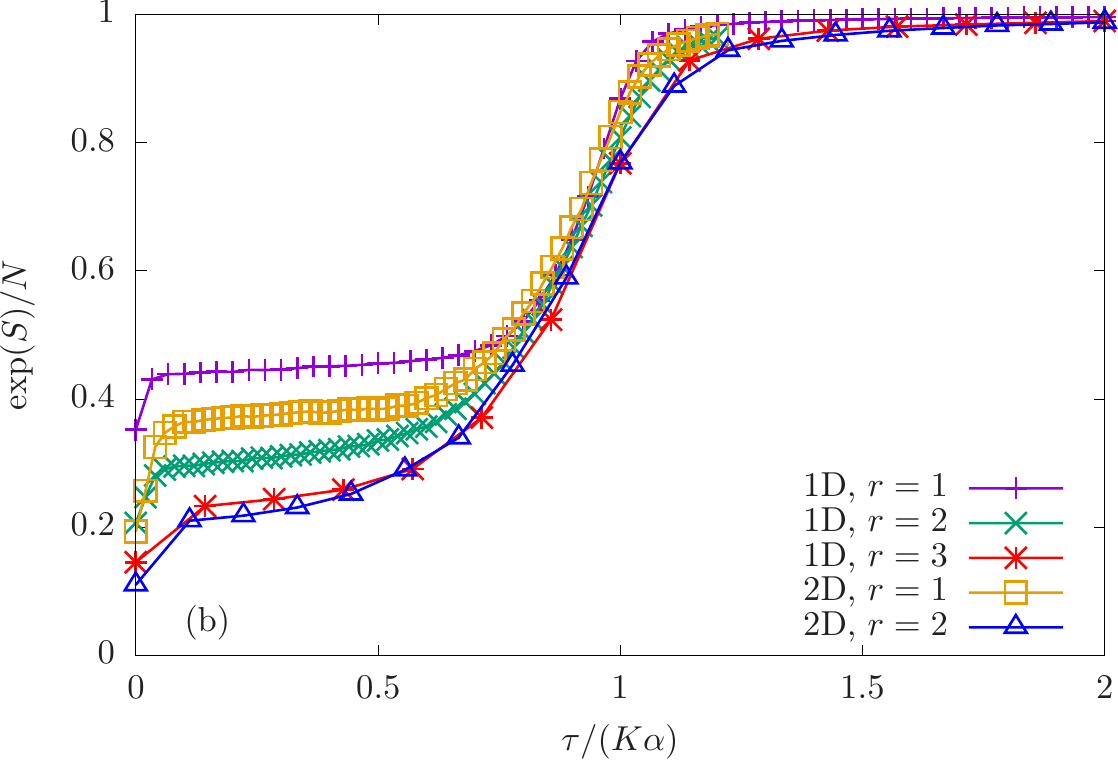}
  \caption{(a) Evolution of the expected value of the fraction of
    effective players, $\exp(S)/N$, as a function of rescaled-time,
    $K\alpha t$, where $K$ is the number of players for each
    round. Notice that, in the long run, the stationary fraction is
    always $1/K$. (b) Expected value of the fraction of effective
    players as a function of $\tau/(K\alpha)$ for different
    topologies. Notice the rather sharp increase for $\tau \sim
    K\alpha$, denoting the phase transition from an unequal to an
    egalitarian system. 1D systems operate with $N=1000$, 2D systems
    with $L_x\times L_y=30\times 30$, so $N=900$.}
  \label{fig:trans_2d}
\end{figure}

Fig. \ref{fig:trans_2d} (a) shows the expected value of the ratio of
effective players, $\exp(S)/N$, as a function of rescaled time:
$K\alpha t$, where $K$ is the number of players in each round,
i.e. the coordination number plus one. In the present 2D case,
$K=5$. We have plotted this average for several topologies: 1D systems
with range 1, 2 and 3 interactions, and 2D systems with range 1 and 2
(next nearest neighbors, so $K=9$). In all cases, the number of
effective players collapses in the short run, while its long-time
behavior changes in a straightforward way: the ratio of effective
players tends to $1/K$.

The short-time collapse of $\exp(S)/N$ in Fig. \ref{fig:trans_2d} (a)
suggests that the relevant time-scale for the development of
inequality in the power game is $1/(K\alpha)$. Fig. \ref{fig:trans_2d}
(b) provides evidence to support this claim, by plotting the expected
value of the long-term fraction of high class agents, $\exp(S)/N$, as
a function of the ratio $\tau/(K\alpha)$ for the different system
topologies explored in this section. Notice that the transition takes
place for $\tau/(K\alpha) \sim 1$, i.e. when the time-scale needed to
build inequality and the time-scale needed to redistribute roughly
coincide.


\section{Conclusions and further work}

In this article we have researched a random non-strategic {\em power
  game} which may describe some features associated to the accretion
of wealth and power in social systems. Even though the elements on
which the game is based are well known in the literature (random
exchanges leading to inequality and smoothing mechanisms), the extreme
simplicity of the model makes it specially interesting and suitable
for analysis in econophysics and in statistical mechanics.

The resource distributed among the agents is called {\em power}
instead of wealth to emphasize its active character: obtaining power
makes an agent more likely to obtaining even more power. In other
words, the probability of winning each round is proportional to the
current power in possession of each agent. Thus, small fluctuations
are naturally enhanced, leading to a morphological instability. At
later stages, the agents can be divided into classes: a high class,
which can call all bets placed by other agents, and a lower class,
which can not. Moreover, the system presents broken ergodicity: there
are many possible stationary states, and an arbitrarily long waiting
time is required to jump from one to another.

Importantly, the game is {\em local}: each agent can only play with
her immediate neighbors. Thus, no two high class players can be
neighbors in the stationary state, since one of them would ruin the
other in the long run. When the players are arranged around a ring,
the dynamics leads to stationary states where the number of effective
players is, in average, one third of the total number of agents. For
other topologies, we found evidence that the final ratio of
low-to-high class players corresponds to the number of players in each
neighborhood, $K$.

Starting from an equal distribution of power, early-time dynamics is
described by Family-Vicsek scaling corresponding to random
deposition. Yet, for longer times, the width of the interface grows
exponentially, thus abandoning the Family-Vicsek paradigm. Shadowing
instabilities typically lead to diffusion limited aggregation (DLA)
behavior \cite{Yao.93,Krug.96,Santalla.18}, which seems to be absent
from our case.

The introduction of a redistribution mechanism can revert this trend
to the growth of inequality, but only if the taxation level is high
enough. We can consider this effect in terms of a competition between
time scales: $\tau^{-1}$ being the time scale in which redistribution
acts and $(K\alpha)^{-1}$, the time scale associated to the natural
growth of inequality. According to our numerical analysis, $\tau >
K\alpha$ leads to an egalitarian society, with restored ergodicity,
while $\tau < K\alpha$ leads to a sharp class division. This result is
reminiscent of Piketty's claim that inequality will grow when the rate
of return of capital $r$ is larger than the rate of growth $g$ of the
whole economy, since $r^{-1}$ and $g^{-1}$ can be thought as the
time scales associated to capital and economic growth.
\cite{Piketty14,Benisty,Berman17b}.

The above behaviours have been analyzed by using different measures
for the inequality of power, such as the roughness of the profile, the
Shannon entropy (which leads to a natural estimate to the number of
effective players) and the Gini coefficient, always leading to similar
conclusions.

Several elements have been left out in this first analysis of the {\em
  power game}. The first one is the characterization of the observed
transition between a class society and an egalitarian society. A
second one is the analysis of more realistic social network
structures. In our 1D and 2D examples, we obtain evidence of a
well-known fact: inequality grows with globalization. In a fragmented
world, an agent can become the richest of a fixed domain, while in a
fully connected world the same agent can become the wealthiest of the
whole system. It is relevant to ask how our observations will change
for strongly inhomogeneous lattices, more similar to human societies,
such as scale-free and small-world networks.

Other interesting element to be considered in future work is the
inclusion of progressive taxation, in which each player $i$ would
contribute a magnitude proportional to $w_i^\beta$, leading to
progressive taxation for $\beta>1$ and regressive for $\beta<1$, which
will present a richer phase diagram. The role of economic initiative
also leads to interesting questions: in principle, it is always
against the interests of low-class players to place a bet. Thus,
low-class players will only play either under coercion or under
incomplete information. Coercion can be modulated through inequality
itself: powerful agents may be able to force the acceptance of the
bet. Yet, the role of incomplete information is specially intriguing,
since it has been empirically found that a large fraction of
respondents have incorrect beliefs about the inequality levels of
their own country \cite{Hauser}. All these possibilities are worth
exploring in future work.


\begin{acknowledgments}
We thank M. Jiménez-Martín, I. Rodríguez-Laguna and
J. Fernández-Pendás for useful conversations. We acknowledge financial
support from the Spanish Government through Grants
FIS2015-69167-C2-1-P, FIS2015-66020-C2-1-P and PGC2018-094763-B-I00.
\end{acknowledgments}


\newpage


\appendix

\begin{widetext}

\section{Alternative set of rules for the two-player power game}
\label{sec:appendix}

As a check, we have considered an alternative version of the power
game, in which the bet is fixed to $\alpha/2$ instead of being
proportional to the power of the gambler. Table \ref{table:fixedbet}
shows the payoff matrix in this case for two players.

\begin{table}[htbp]
\centering
\begin{tabular}{|c|c||c|c|}
  \hline
 Gambler & Winner & Probability & Payoff \\
 \hline \hline
 $1$ & $1$ & $\frac{1}{2} w_1$ & $+\alpha/2$\\
\hline
 $1$ & $2$ & $\frac{1}{2} (1-w_1)$ & $-\alpha/2$\\
 \hline
 $2$ & $1$ & $\frac{1}{2} w_1$ & $+\alpha/2$\\
\hline
 $2$ & $2$ & $\frac{1}{2} (1-w_1)$ & $-\alpha/2$\\
\hline
\end{tabular}
\caption{Alternate power game, with a fixed bet.}
\label{table:fixedbet}
\end{table}

Please notice that, in this variant of the game, the choice of the
gambler is irrelevant, since the bet will always be the same.  We can
estimate in this case the expected payoff and its variance, as we did in
the standard power game, obtaining the following result:

\begin{eqnarray}
 \textrm{E}( \textrm{Payoff}) &=& \alpha \(w_1-\frac{1}{2}\), \label{eq:fbmean}\\
 \textrm{V}( \textrm{Payoff}) &=& \alpha^2 w_1 \(1-w_1\). \label{eq:fbdev}
\end{eqnarray}

We observe that the expected gain coincides with the standard game,
but the deviation is proportional to the initial value of the power of
player 1. Fig. \ref{fig:juego2} shows the histograms, expected power
and deviation of the power of player 1, similarly to
Fig. \ref{fig:juego1}. One can see that the histogram only contains
delta peaks, since the amounts of power of player 1 can only be given
by $w_1(t)=w_1(0) + m\alpha$, where $m\in \Z$.

\begin{figure}
  \hbox to 16cm{
  \includegraphics[width=6.5cm]{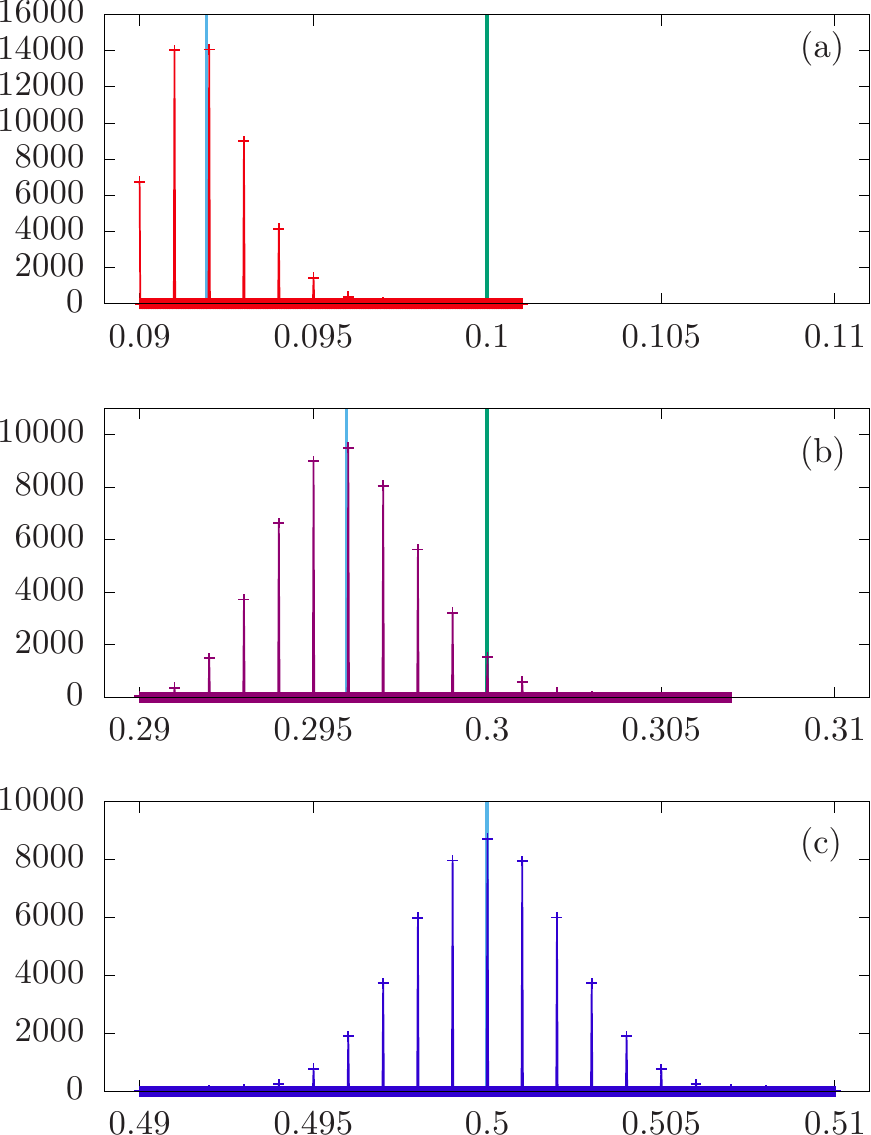}
  \includegraphics[width=7cm]{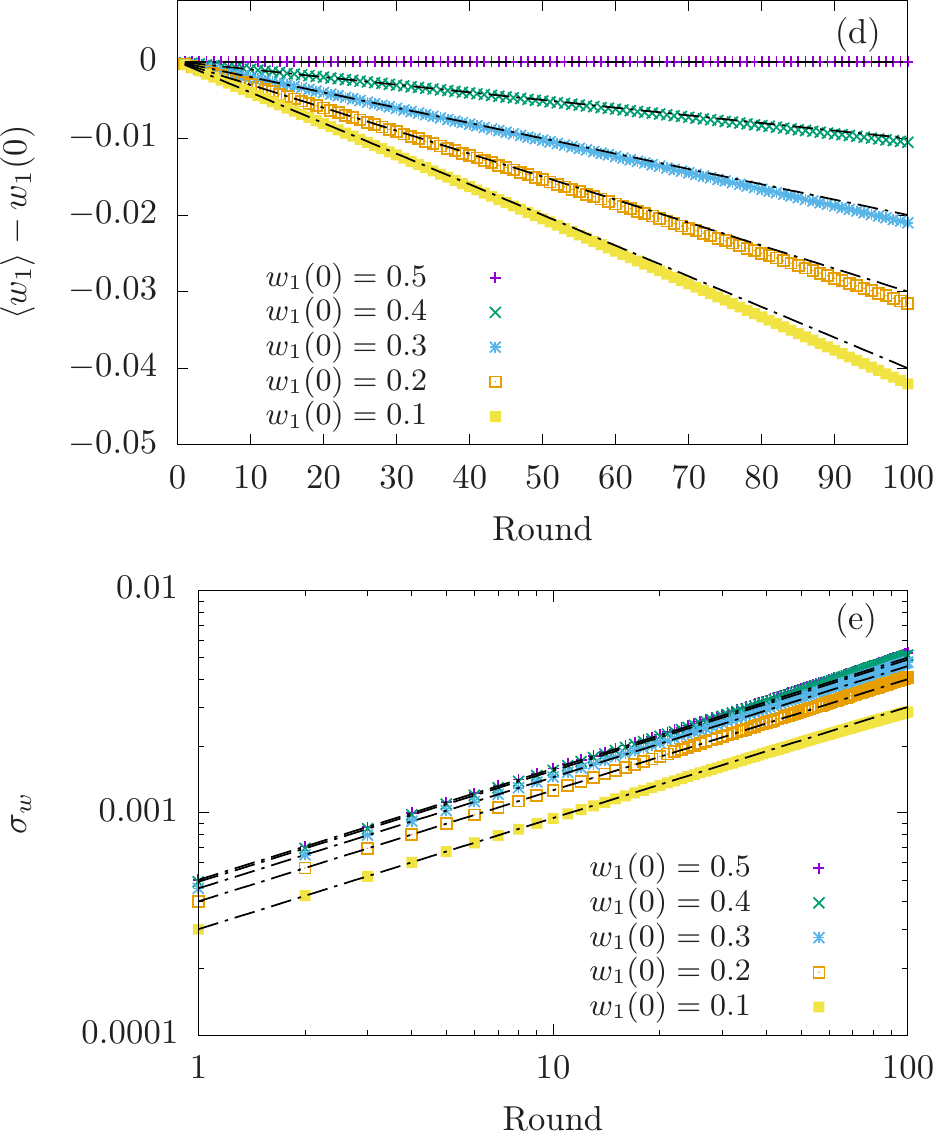}}
  \caption{Alternate power game, with fixed bet $\alpha/2$. Left:
    two-player histogram of $w_1(t)$ using $\alpha=10^{-3}$ for
    different initial values $w_1(0)=0.1$ (a), $0.3$ (b) and $0.5$ (c)
    and 20 games, using $10^6$ realizations for each case. (d) Average
    power of player 1 for the different initial conditions,
    $w_1(0)=0.1$ up to $0.5$ as a function of the round order. (e)
    Power deviation for the same cases. Dashed lines constitute
    theoretical predictions, Eqs. \eqref{eq:fbmean} and (the square
    root of) \eqref{eq:fbdev}.}
  \label{fig:juego2}
\end{figure}

\end{widetext}

\end{document}